\documentclass[journal=ancac3,manuscript=article]{achemso}

\usepackage{chemformula} % Formula subscripts using \ch{}
\usepackage[T1]{fontenc} % Use modern font encodings

\author{Yilan Wang}
\affiliation
{School of Microelectronics, South China University of Technology, Guangzhou, 510641, China}
\author{Feng Tian}
\affiliation
{School of Microelectronics, South China University of Technology, Guangzhou, 510641, China}
\author{Wendi Huang}
\affiliation
{School of Microelectronics, South China University of Technology, Guangzhou, 510641, China}
\author{Taojie Zhou}
\affiliation
{School of Microelectronics, South China University of Technology, Guangzhou, 510641, China}
\email{taojiezhou@scut.edu.cn}
\title[An \textsf{achemso} demo]
  {Large-angle twisted photonic crystal semiconductor nanolasers with ultra-low thresholds operating in the C-band}


\begin{document}


%\begin{tocentry}

%\includegraphics[width=\textwidth]{TOC}

%	Some journals require a graphical entry for the Table of Contents.
%	This should be laid out ``print ready'' so that the sizing of the
%	text is correct.
%	
%	Inside the \texttt{tocentry} environment, the font used is Helvetica
%	8\,pt, as required by \emph{Journal of the American Chemical
%		Society}.
%	
%	The surrounding frame is 9\,cm by 3.5\,cm, which is the maximum
%	permitted for  \emph{Journal of the American Chemical Society}
%	graphical table of content entries. The box will not resize if the
%	content is too big: instead it will overflow the edge of the box.
%	
%	This box and the associated title will always be printed on a
%	separate page at the end of the document.
	
%\end{tocentry}

\begin{abstract}
Nanolasers, characterized by enhanced optical localization at subwavelength scale, have emerged as promising coherent light sources for ultra-compact, high-speed and energy-efficient photonic integrated circuits. Twisted photonic crystal nanocavity, constructed by stacking two layers of photonic crystal structure with a specified rotation angle, enables strong light confinement with an ultra-small mode volume and an extremely high quality factor. The twisted angle can be randomly selected, providing the possibility of actively tuning the resonant wavelength and optical mode distribution within a nanoscale twisted cavity. Here, we demonstrate large-angle twisted single-mode photonic crystal nanolasers operating in the C-band with an exceptionally ultra-compact footprint of approximately 25 $\mu m^2$ and an ultra-small mode volume of 0.47 $(\lambda/n)^3$. The reported twisted photonic crystal nanolasers are optically pumped at room temperature with an ultra-low threshold of $\sim$ 1.25 $kW/cm^2$. Our work provides a prospective method for easily constructing robust nanolasers by twisting angles, and paves the way for achieving high-performance nanoscale coherent light sources for densely integrated photonic chips.
\end{abstract}

\section{Introduction}
Photonic integrated circuits (PICs), utilizing light as a medium for information transfer, are regarded as a crucial foundation for the next-generation information technology with high energy-efficient and high-throughput data processing\cite{nagarajan2005large,komljenovic2015heterogeneous,dong2014silicon}. However, the performance of PICs heavily depends on the efficiency and stability of their core components, particularly semiconductor lasers responsible for signal generation\cite{zhou2023prospects,zhou2015chip}. As PICs continue to advance, there is an increasing demand for laser technology that enables further device miniaturization, improved energy efficiency, and wavelength tunability\cite{chen2011nanolasers,sarkar2023ultrasmall,deka2020nanolaser}. In this context, semiconductor nanolasers, distinguished by their sub-wavelength dimensions, offer significant advantages in terms of dense integration, low power consumption and fast modulation speed\cite{ma2019applications,ning2019semiconductor,matsuo201120,takeda2013few}. These attributes make them highly promising for diverse applications\cite{liang2020plasmonic,ma2019applications}, including on-chip ultra-compact coherent light sources for PICs\cite{zhou2020continuous,chen2011nanolasers}.

Inspired by the intriguing features of twisted van der Waals materials\cite{he2021moire,yankowitz2019tuning,lisi2021observation}, Moiré superlattice nanophotonic devices, especially Moiré lasers, have recently emerged as a rapidly growing research field due to their unique optical properties such as flat band dispersion\cite{du2023moire,sunku2018photonic,talukdar2022moire}. However, the generation of a Moiré photonic flat band typically relies on a considerably long-length scale photonic crystal (PhC) structure\cite{tang2020photonic,mao2021magic,luan2023reconfigurable,raun2023gan}, contrary to the trend of device miniaturization. To overcome the spatial limitation of Moiré lattices, an ultra-compact twisted PhC nanocavity utilizing a limited number of PhC periods has recently been developed\cite{ma2023twisted,ouyang2024singular}, constructed by truncating a segment of the Moiré PhC lattices. Unlike traditional PhC defect nanocavities\cite{akahane2003high,zhou2021single,song2019ultrahigh}, which generally require carefully optimized airhole distributions to achieve a high quality factor (Q-factor) and a small mode volume ($V_{m}$), the twisted PhC nanocavity accommodates for arbitrary twist angles and enables strong light confinement with both a high Q-factor and an ultra-small $V_{m}$\cite{ma2023twisted}. Thus, the twisted nanocavity allowing easy configuration provides a powerful scheme for designing nanoscale lasers with ultra-low threshold and small footprint. Despite these promising characteristics, high-performance semiconductor nanolasers in the telecom band utilizing ultra-compact large-angle twisted PhC nanocavities have yet to be fully explored, limiting their potential applications in high-density photonic chips.

Here, we present ultra-compact large-angle twisted PhC semiconductor nanolasers with ultra-low lasing thresholds in the C-band, utilizing InGaAsP multi-quantum wells as the gain medium. The large-angled (5$^{\circ}$) twisted PhC nanolasers, featured with an ultra-small footprint of approximately 25 $\mu m^2$ and a theoretical ultra-small $V_{m}$ of $\sim$ 0.47 $(\lambda/n)^3$, are optically pumped at room temperature with single-mode emission and an exceptionally low lasing threshold of $\sim$ 1.25 $kW/cm^2$. Meanwhile, the temperature-dependent operation of the twisted nanolaser is also performed to see their temperature stability. The twisted photonic crystal lasers presented in this work offer significant advantages in terms of low threshold and compact size, and hold great promise as nanoscale coherent light sources for densely integrated photonic chips.

\section{Results and discussion}

%\subsection{Simulation}

\begin{figure}[htb]
	\centering
	\includegraphics[width=1\textwidth]{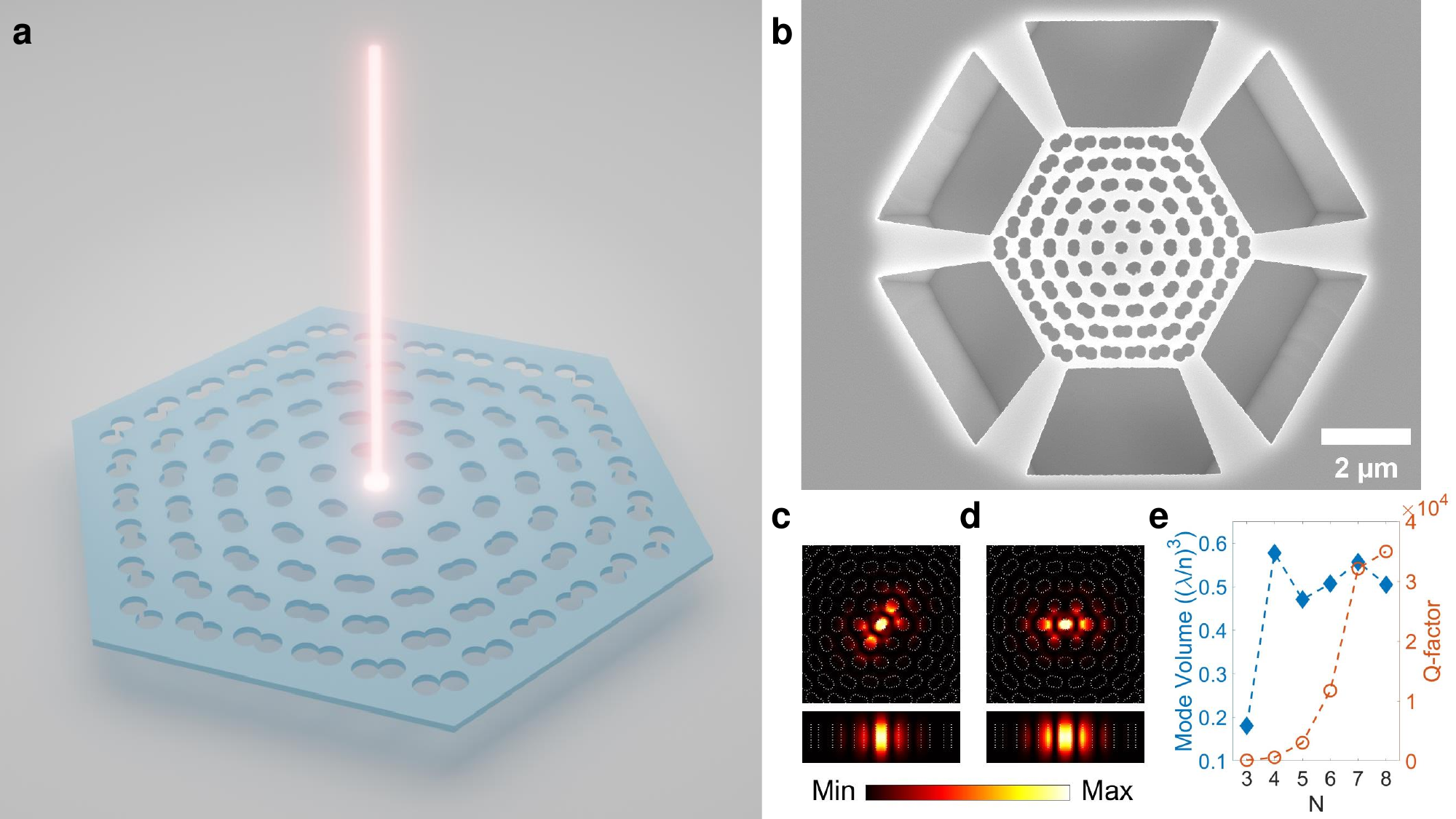}
	\caption{(a) Schematic diagram of a twisted PhC nanolaser. (b) Top-view SEM image of the fabricated twisted PhC nanolaser. (c-d) Calculated top (top panel) and cross-sectional (bottom panel) views of E-field profiles for the doubly degenerated modes in the twisted PhC nanocavity. (e) The simulated $V_{m}$ and Q-factor under various periods of PhC ($N$).}
	\label{1}
\end{figure}

A schematic diagram of the large-angle twisted PhC nanolaser is illustrated in Fig. 1a, showing a strong light confinement in the central region for laser emission. Figure 1b presents a scanning electron microscope (SEM) image of the fabricated twisted PhC nanolaser. Six trapezoidal structures around the twisted PhC region are incorporated to enhance the wet-etching process and provide additional mechanical support for the suspended membrane. Figures 1c and 1d show the top and cross-sectional views of the calculated electric field profiles for the doubly degenerated high-Q fundamental modes in a twisted nanocavity with lattice constant $a =$ 540 nm, airhole radius $r/a =$ 0.26, and a twisted angle of \(5^\circ\). The top-view electric field profiles reveal a notable \(60^\circ\) deviation in the polarization directions of the doubly degenerated optical modes, attributed to the rotational symmetry inherent in the twisted PhC nanocavity. The $V_{m}$ and Q-factor of the fundamental resonant mode for various numbers of PhC periods (\(N\)) are shown in Fig. 1e, showing a rapidly increase of Q-factor and a nearly saturated $V_{m}$ as increasing the periods $N$. Here, we choose $N = 5$ for the designed structure with a relatively high Q-factor of approximately $3100$ and an ultra-small $V_{m}$ of 0.47 $(\lambda/n)^3$.

%\subsection{Optical characterisation}

\begin{figure}[htb]
	\centering
	\includegraphics[width=1\textwidth]{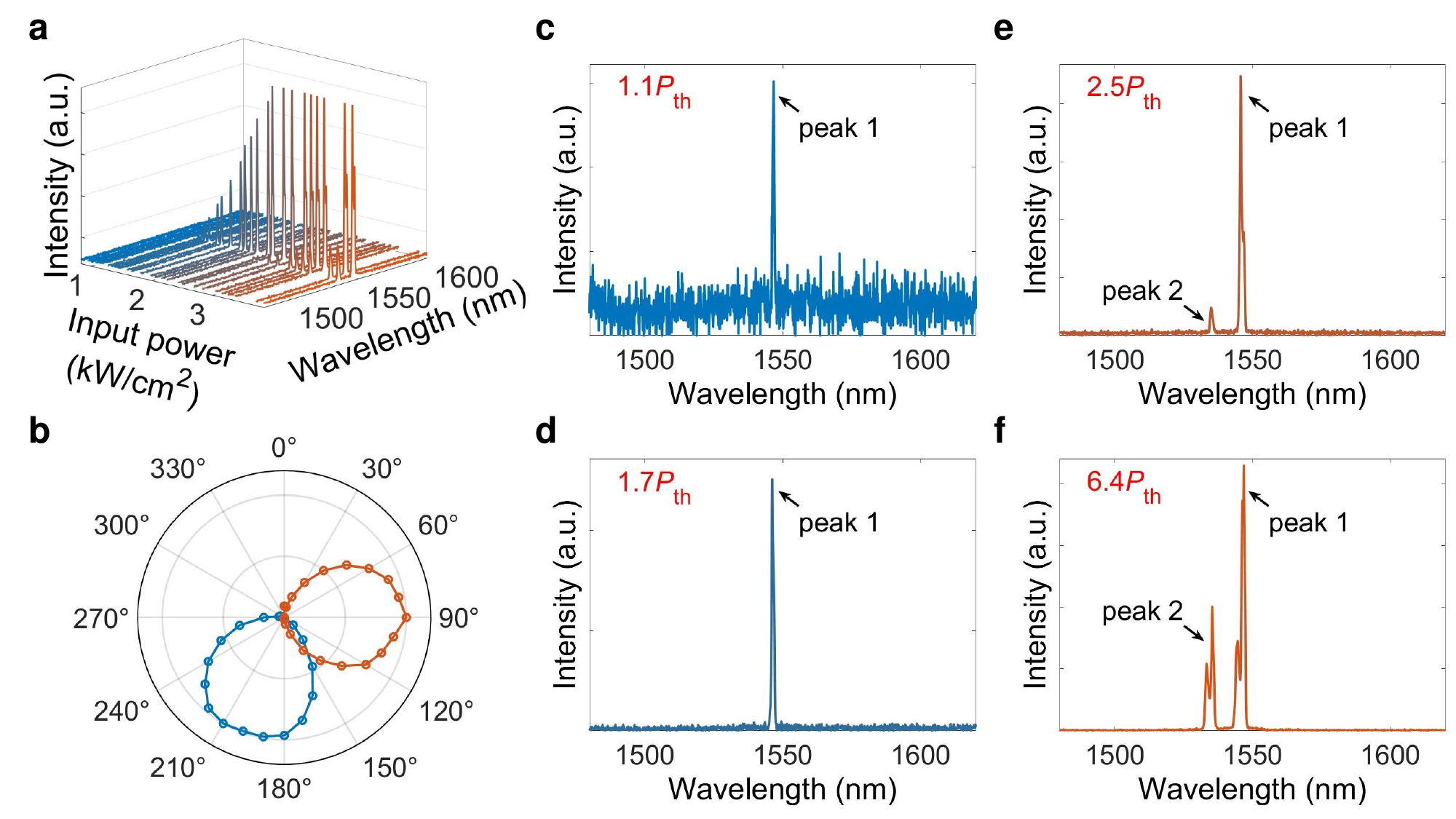}
	\caption{(a) Measured power-dependent spectra of a twisted PhC nanolaser with $a =$ 540 nm and $r = 0.26a$. (b) Normalized polarization-dependent output intensity of the fundamental (blue, 1547 nm) and higher-order (orange, 1535 nm) lasing modes. Measured lasing spectra under an input power of 1.1$P_{th}$(c), 1.7$P_{th}$(d), 2.5$P_{th}$(e) and 6.4$P_{th}$(f).}
	\label{2}
\end{figure}

The ultra-compact large-angle twisted PhC nanolasers with InGaAsP multi-quantum wells as gain materials are fabricated and optically pumped at room temperature in a micro-photoluminescence ($\mu$-PL) system. Figure 2a shows the measured power-dependent spectra of a twisted PhC nanolaser with $a =$ 540 nm and $r/a =$ 0.26. A distinct high-Q factor fundamental mode (peak1, 1547 nm) emerges from the background of spontaneous emission and transits to stimulated radiation as increasing the input powers, and a side peak with a shorter wavelength (peak2, 1535 nm) appears at elevated input powers due to the mode competitions. The polarization of the two lasing peaks is measured by putting a polarizer in front of the spectrometer as shown in Fig. 2b. The correlation between output intensities and polarization angles at 1547 nm (peak1) and 1535 nm (peak2) are respectively presented in blue and orange curves, showing the linear polarization emission for the two lasing peaks. Notably, due to the rotational symmetry, the polarization angles corresponding to the two peaks differ by \(60^\circ\). Figures 2c$-$2f display the measured spectra at varying input powers. The twisted nanolaser exhibits a singular sharp lasing peak without obvious mode splitting below 2.5$P_{th}$. As increasing the input powers, each degenerated peaks are split into two non-degenerated modes owing to the fabrication imperfections.

\begin{figure}[t!]
	\centering
	\includegraphics[width=1\textwidth]{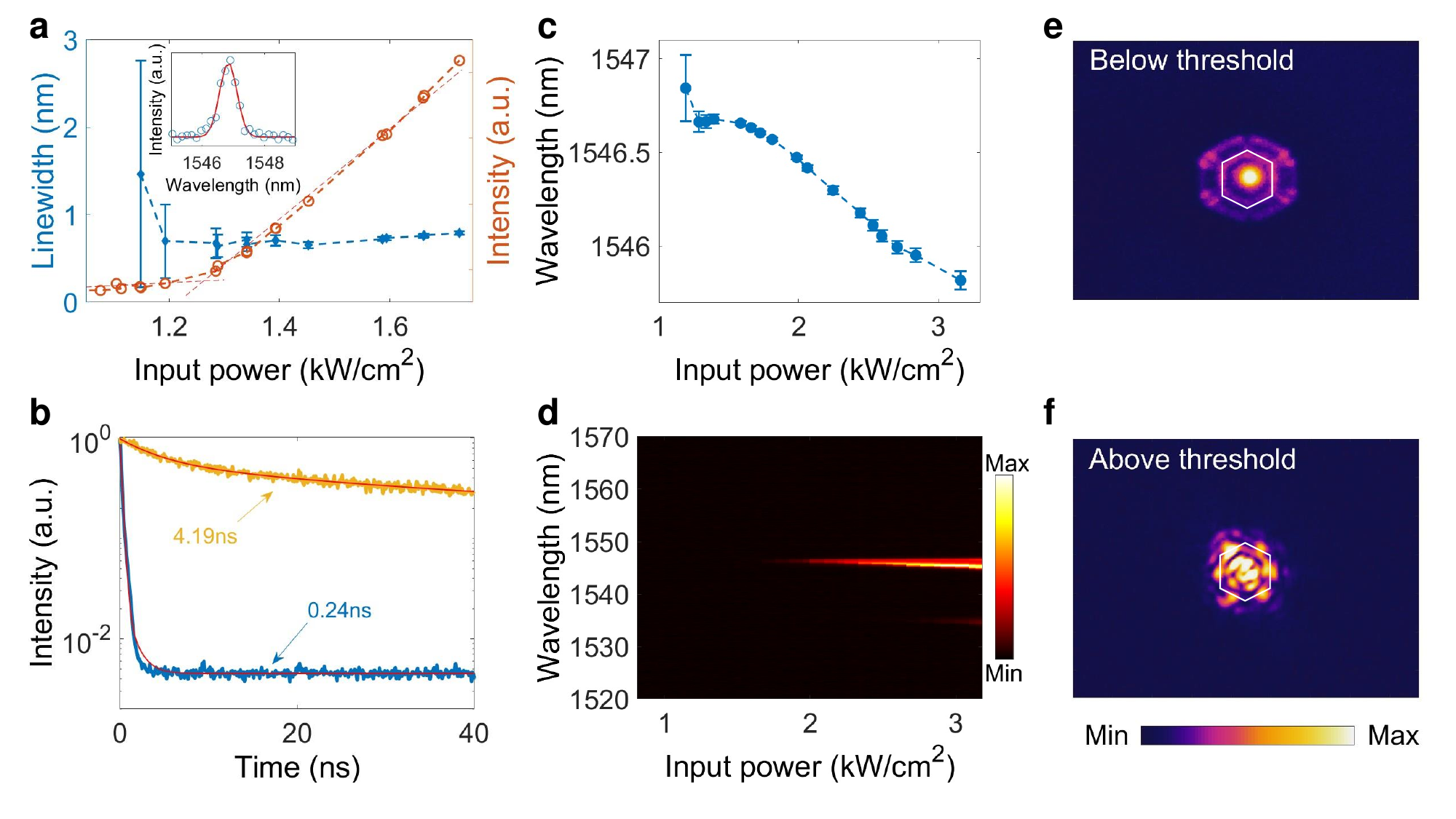}
	\caption{(a) Collected $L$-$L$ curve and linewidth of the lasing peak at 1547 nm, indicating a lasing threshold of $\sim$ 1.25 $kW/cm^2$. The inset shows curve fitting (red line) of measured data (open circles) just below the threshold. (b) Collected normalized TRPL spectra of spontaneous emission (orange curve) and stimulated emission (blue curve). The red curves are fits to a bi-exponential decay model. Measured lasing wavelengths (c) and image of laser spectra (d) under various input powers. The near field profiles measured below (e) and above (f) the threshold. The white line presents the boundary of the fabricated twisted nanolaser.}
	\label{3}
\end{figure}

Figure 3a illustrates the collected light-in/light-out ($L$-$L$) curve and the linewidth of the lasing peak at 1547 nm with an estimated threshold of approximately 1.25 $kW/cm^2$, providing clear evidences of lasing operation through a kink in the $L$-$L$ curve and the linewidth narrowing effect. The inset of Fig. 3a displays a Gaussian curve fit (solid curve) to the experimental spectra collected near the threshold, indicating a linewidth of approximately 0.67 nm and an experimental Q-factor $\sim$ 2300 (\(Q = \lambda/{\Delta\lambda}\), where \(\lambda\) signifies the resonance frequency of the cavity, and \(\Delta\lambda\) denotes the linewidth of the resonance peak). The dynamic behavior of the fabricated twisted nanolaser is further estimated by conducting the time-resolved PL (TRPL) measurement. Figure 3b depicts the normalized TRPL spectra for the unpatterned region (orange curve) and the twisted nanolaser (blue curve) collected above the threshold, exhibiting the typical exponential decay characteristic and a clear transition from spontaneous emission ($\tau_{spe} =$ 4.19 ns) to the stimulated radiation with a shorter lifetime ($\tau_{lasing} =$ 0.2 ns). The lasing wavelengths of the fundamental mode and normalized spectra measured under various input powers are displayed in Fig. 3c and Fig. 3d, respectively. A blue-shift in the lasing wavelength is observed after the threshold, primarily attributed to the band-filling effect\cite{mylnikov2020lasing} without evident heat accumulation due to the pulse pumping conditions. Figures 3e and 3f present the near-field intensity distribution of the lasing peak at 1547 nm, captured by using an InGaAs camera. Prior to reaching the threshold, the twisted PhC nanocavity is dominated by spontaneous emission. Strong speckle patterns appear around the device after the threshold, owing to the highly coherent emission.

\begin{figure}[t]
	\centering
	\includegraphics[width=1\textwidth]{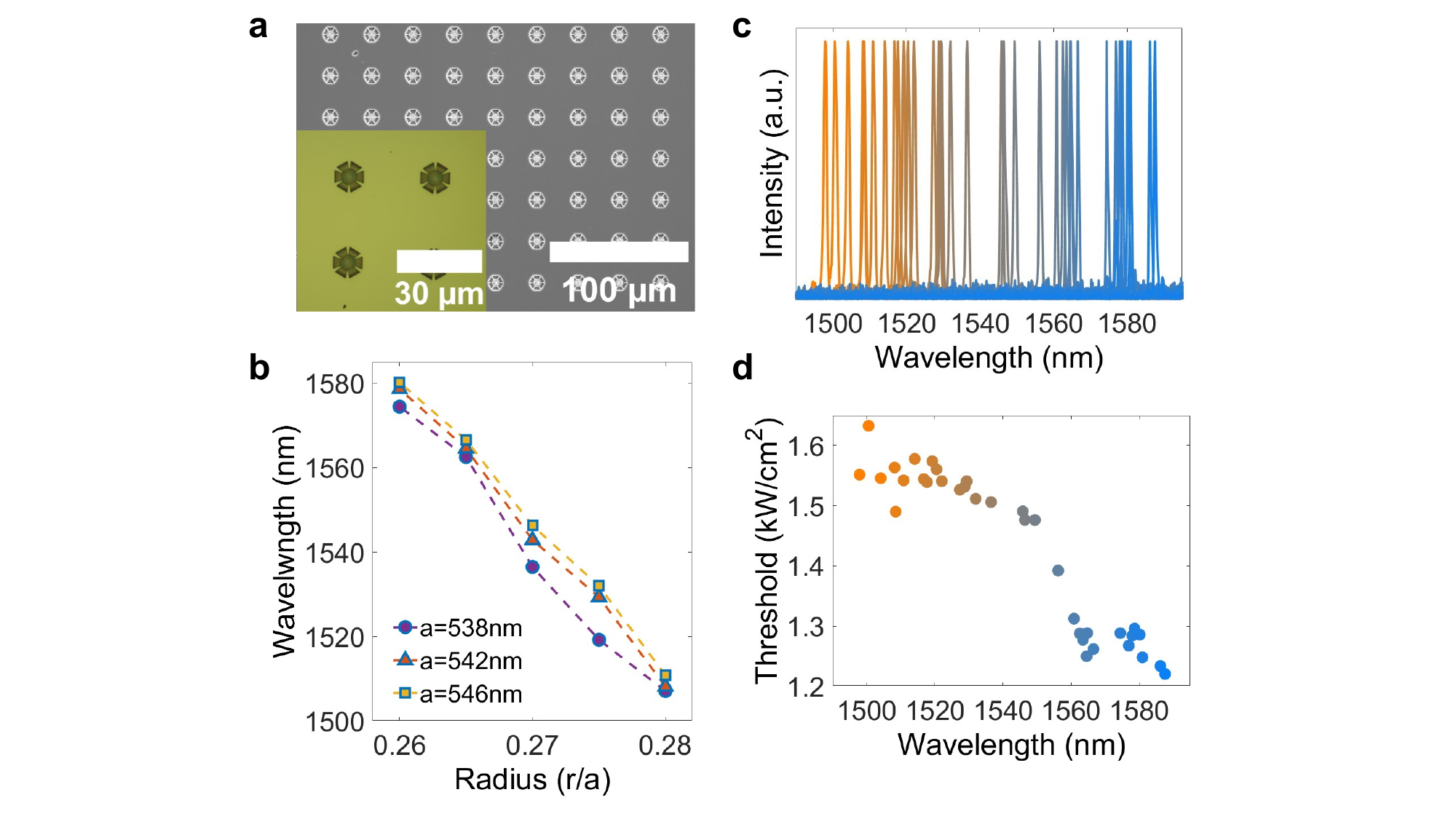}
	\caption{(a) Top view SEM image of an array of ultra-compact twisted PhC nanolasers. The inset shows an optical microscopy image of fabricated devices. (b) The lasing wavelengths under various lattice constant ($a$) and radius ($r/a$). (c) Normalized measured lasing spectra from representative twisted PhC nanolasers with various structural parameters, showing a tunable range of approximately 85 nm. (d) The corresponding measured thresholds for various lasing peaks.}
	\label{4}
\end{figure}

High density integrated multi-wavelength light sources are crucial for energy-efficient and high-throughput photonic integrated circuits such as on-chip wavelength-division-multiplexing systems\cite{dumont2022high}. Here, by manipulating the lattice constant ($a$) and the radius of the airhole ($r$), we present multi-wavelength emission from an array of single-mode twisted PhC nanolasers. A SEM image of fabricated array with various structural parameters ($a$ and $r$) is shown in Fig .4a, and the inset presents an optical microscope image of the fabricated devices. The lasing wavelengths can be finely tuned by precisely control the $a$ and $r$. Figure 4b presents the lasing wavelengths of twisted nanolasers with various radii of airholes ($r/a$) from 0.26 to 0.28 for lattice constant $a =$ 538 nm, 542 nm, and 546 nm, respectively. Figure 4c shows the normalized lasing spectra from the fabricated single-mode ultra-compact twisted nanolasers, indicating a wide tunable wavelength range of approximately 80 nm in the 1550 nm telecom band. The corresponding distribution of experimental lasing thresholds is shown in Fig. 4d.  A slightly decrease of threshold at longer wavelength is observed, suggesting a higher gain at long wavelength region. The laser wavelengths exhibit a gradually linear red-shift for larger $a$ and smaller $r$ owing to the increase of effective refractive index, agreeing well with the simulated trend of changes in resonant peaks for the designed large-angle twisted PhC nanocavity. This precise control over the wavelength through tuning the geometrical parameters renders the large-angled twisted nanolasers suitable for high density integrated photonic chips.

\begin{figure}[t]
	\centering
	\includegraphics[width=1\textwidth]{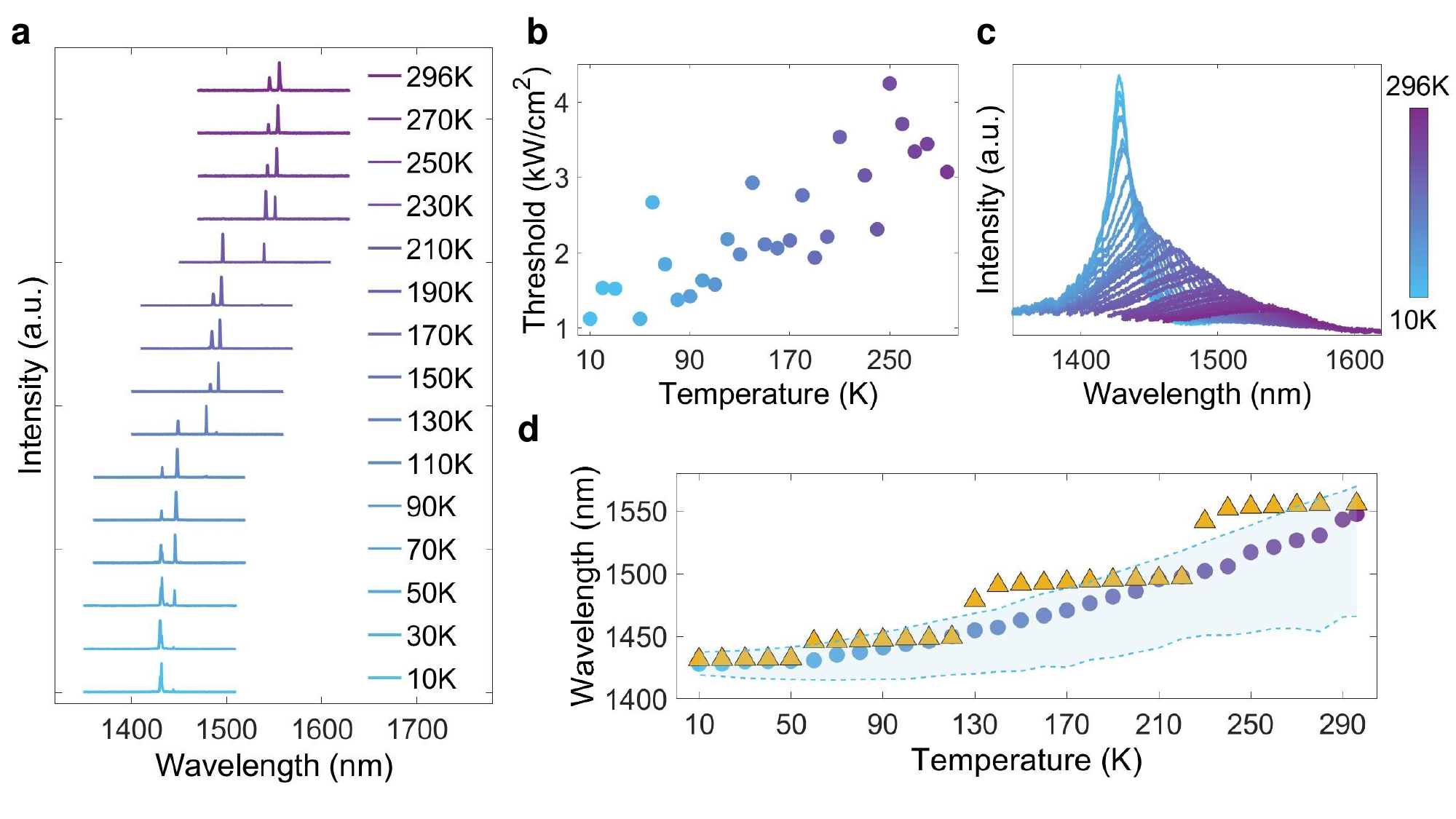}
	\caption{(a) Temperature-dependent lasing spectra measured from 10 to 296 K for a large-angle twisted PhC nanolaser with $a =$ 540 nm and $r/a =$ 0.26. (b) The distribution of lasing thresholds under various temperatures. (c) Measured temperature-dependent spontaneous emission spectra. (d) The lasing peaks (triangles), central wavelengths (circles) and FWHM (blue filled region) of spontaneous emission under various temperatures.}
	\label{5}
\end{figure}

Furthermore, the temperature-dependent lasing property of the fabricated twisted nanolaser is performed from 10 K to room temperature by mounting the device in a liquid helium cryostat. Figure 5a shows the measured lasing spectra at elevated input powers measured under various temperatures for a fabricated twisted nanolaser with $a =$ 540 nm and $r/a =$ 0.26. Multi-mode emission is observed under high input powers and the lasing wavelength is red-shift as increasing the temperature owing to the temperature-dependent gain spectra and resonant frequencies. The nonradiative recombination rate of the fabricated twisted PhC nanolasers with high surface-to-volume ratio is expected to be dramatically suppressed as decreasing the temperature, resulting in higher gain and reduced lasing threshold. The threshold of the main lasing peak at various temperatures is shown in Fig. 5b, indicating a trend of reduction from nearly 4 $kW/cm^2$ at high temperature to 1 $kW/cm^2$ at liquid helium temperature. Meanwhile, a slight vibration of lasing thresholds exists as varying the temperature. The experimental threshold at 250 K ($P_{th} \sim$ 4 $kW/cm^2$) is larger than that of room temperature ($P_{th} \sim$ 3 $kW/cm^2$), which is different from the usual tendency. To further understand the discrepancy, the wavelengths of the lasing peak are compared with the gain spectra estimated from spontaneous emission. Figure 5c presents the measured spontaneous spectra from an unpatterned area, showing a narrower full-width half-maximum (FWHM) and higher output intensity at lower temperature. The corresponding central wavelengths (circles) and FWHM (blue filled regions) of spontaneous spectra at various temperatures are shown in Fig. 5d, with marked triangles indicating the lasing peaks. A slightly increased mismatch between the lasing peak and gain spectra is observed for temperature ranging from 296 K to 250 K, leading to lower gain and higher lasing threshold.

\section{Conclusions}

In summary, we propose a simple method to construct ultra-compact PhC nanocavities by twisting large angle between two sets of PhC structures. Single-mode large-angle twisted PhC nanolaser with an ultra-small footprint of $\sim$ 25 $\mu m^{2}$ and a $V_{m}$ of 0.47 $(\lambda/n)^{3}$ is demonstrated at room temperature. The lasing operation is confirmed by the collected $L$-$L$ curve, linewidth narrowing and near-field mode profile. Mode competition is observed under elevated input powers with high-order resonant mode and doubly degenerated mode splitting. By precisely tuning the structural parameters ($a$ and $r$), multi-wavelength single-mode emission is achieved in an array of fabricated devices with a broad output range of approximately 85 nm within the 1550 nm telecom band. The temperature-dependent characteristics of the twisted PhC nanolaser are performed, showing stable lasing emission from 10 K to 296 K with a blue-shift of the lasing wavelengths as decreasing the temperature. Our work provides a promising method for easily constructing robust nanolasers by twisting angles, and paves the way for achieving high-performance nanoscale coherent light sources for densely integrated photonic chips.

\section{Methods}

\subsection{Simulation} Three-dimensional finite-difference time-domain method is implemented to simulate the resonant frequency, Q-factor and $V_{m}$ of the twisted photonic crystal nanocavity. The mode volume is calculated using the formula \(V_{\text{mode}} = \frac{\int \epsilon(\mathbf{r}) \mathbf{E}(\mathbf{r}) ^2 \, d\mathbf{r}}{max(\epsilon(\mathbf{r}) \mathbf{E}(\mathbf{r})^2)}\), where \(\epsilon(\mathbf{r})\) denotes the dielectric constant as a function of spatial position, \(\mathbf{E}(\mathbf{r})\) represents the electric field intensity at a given position.

\subsection{Fabrication}
The twisted PhC nanolasers were fabricated based on a 260-nm-thick slab containing six strained InGaAsP multi-quantum wells. Electron beam lithography was employed to define the designed pattern. The pattern was subsequently transferred through a 90-nm-thick SiO$_{2}$ hard mask and into the active layer by plasma dry etching. A mixture of Cl$_{2}$ and N$_{2}$ was conducted to achieve vertical and smooth etching profiles. Then, the remaining SiO$_{2}$ was removed in a buffered oxide solution. Finally, diluted hydrochloric solution is utilized to etch the 1.3-$\mu m$-thick InP sacrificial layer, forming a suspended PhC membrane.

\subsection{Optical Measurement}
The twisted PhC nanolasers were optically pumped using a 632 nm pulsed laser (10 ns, 800 kHz) in a micro-photoluminescence system. The pumping spot with a diameter of approximately 2 $\mu m$ was focused with a 100$\times$ objective lens and precisely positioned at the nanocavity through piezo-electric nanopositioners. The emission spectra were collected from the top through the same objective lens and analyzed by the spectrometer with a liquid nitrogen-cooled infrared InGaAs detector. The near-field images were collected by using an InGaAs camera. The carrier lifetime measurements were conducted in a TCSPC system, and the carrier lifetimes were extracted from fitting the experimental spectra with a standard bi-exponential component function ($\textit{I}(\textit{t}) = $\textit{A}$_{1}$exp(-\textit{t}/$\tau$$_{1}$)$ +$ \textit{A}$_{2}$exp(-\textit{t}/$\tau$$_{2}$)), where $\tau$$_{1}$ and $\tau$$_{2}$ are the fast and slow decay components, respectively. For temperature-dependent measurements, the fabricated twisted PhC nanolasers were mounted in a liquid helium cryostat with the temperature controlled from 10 K to 296 K.

\begin{acknowledgement}

 We acknowledge the financial support from the National Natural Science Foundation of China (Grant No. 62304080), the Guangdong Basic and Applied Basic Research Foundation (Grant No. 2024A1515010802), the Science and Technology Projects in Guangzhou (Grant No. 2024A04J3683), the Fundamental Research Funds for the Central Universities (Grant No. 2023ZYGXZR068). T.Z. acknowledges the startup funds from South China University of Technology.

\end{acknowledgement}

\section{Author contributions}

Y.W. carried out the simulation, optical measurement and wrote the paper. F.T. and W.H. fabricated the devices. T.Z. supervised the project.

\section{Competing interests} 
The authors declare no competing interests.

\section{Data availability} 
The data are available from the corresponding author upon reasonable request.

%\begin{suppinfo}
%
%A listing of the contents of each file supplied as Supporting Information
%should be included. For instructions on what should be included in the
%Supporting Information as well as how to prepare this material for
%publications, refer to the journal's Instructions for Authors.
%
%The following files are available free of charge.
%\begin{itemize}
%  \item Filename: brief description
%  \item Filename: brief description
%\end{itemize}
%
%\end{suppinfo}

\bibliography{achemso-demo}

\end{document}